\documentclass[]{krakproc}
\usepackage{graphicx}
\begin{document}

\title{The Neutral Medium}
\author{Robert Braun}
\institute{ASTRON, Postbox 2, 7990 AA Dwingeloo, The Netherlands}
\markboth{R. Braun}{The Neutral Medium}

\maketitle

\begin{abstract}
We consider the physical conditions of the neutral medium within, and in the
environments of, galaxies. The basic physical and morphological properties of
the neutral medium within galaxy disks are now quite
well-constrained. Systematic variations in temperature and phase-balance (of
cool versus warm neutral gas) are indicated as a function of both radius and
$z$-height. Interestingly, the cool medium line-widths are observed to be
dominated by turbulent energy injection within cells of 10~pc to 1~kpc
size. Deep new observations reveal that 5--10\% of the neutral medium is
associated within an extended halo which rotates more slowly and experiences
radial inflow. Much of this component is likely to be associated with a
``galactic fountain'' type of phenomenon. However, compelling evidence is also
accumulating for the importance of tidal disruption of satellites as well as
continuous accretion (of both diffuse and discrete components) in fueling
galaxy halos and disks. Continued fueling is even observed on scales of
100's of kpc in galaxy environments, where the neutral component is likely to
be merely a trace constituent of a highly ionized plasma.
\end{abstract}

\section{Introduction}

Although the neutral interstellar medium, as traced by the 21~cm emission line
of atomic hydrogen, is often imagined to be a rather homogenous environment,
this is really not the case. The neutral medium, essentially by definition (if
we adopt, say, a 50\% or greater neutral fraction), occupies the regime
between the ionized medium on the one hand and the molecular medium on the
other. Within this regime, equilibrium gas kinetic temperatures vary from some
10$^4$~K to as low as 10~K.  Even the boundaries of the neutral regime are
somewhat indistinct, since trace atomic components (as low as 1~\% or less by
number) are often still the most effective observational probes of both highly
ionized plasmas as well as predominantly molecular regions. In this short
review we will consider the entire neutral regime as well as some of these
boundary zones, first from a theoretical perspective and then an observational
one. 

\section{Predicted Thermodynamics, Phase Balance and Temperature}

A critical concept to appreciate is the fact that atomic hydrogen can exist in
pressure equilibrium in two distinct thermodynamic phases, the so-called cool
neutral medium (CNM) and the warm neutral medium (WNM). These are much more
than merely slightly warmer or cooler regions of interstellar space. They are
distinct phases which are separated by a phase transition, that can perhaps
best be visualized by considering the phase transition between liquid water
and water vapor near the earth's surface. A similar density contrast, of
about a factor of 100, is associated with both of these phase transitions.
The concept of an interstellar cloud, should then have somewhat more meaning
to the reader. Although still diffuse by earthly standards, these are much
more substantial than the medium within which they are immersed.

Physical conditions in the neutral medium have been the subject of a series of
analyses going back to the pioneering work of Field (\cite{field65}),
continuing with Draine (\cite{draine78}), Shull \& Woods (\cite{shulland85})
and most recently with Wolfire et al. (\cite{wolfireetal03}). A wide variety
of heating and cooling mechanisms play a role in determining the temperature
and phase balance of the neutral medium, but as soon as even moderate metal
abundances are attained, the heating is dominated by photo-electric emission
from dust. Cooling, on the other hand remains sensitive to the particle
density, with Ly$\alpha$ emission dominating at low densities (below about
1~cm$^{-3}$) and \ion{C}{II} emission at high. Equilibrium solutions have a
characteristic tilted ``S'' shape when plotted in the pressure -- volume
density plane, as seen for example in Fig.~7 of Wolfire et
al. (\cite{wolfireetal03}). Thermodynamically stable solutions are those with
a positive slope in these figures, while the negative slopes correspond to
unstable conditions. Small temperature perturbations will lead parcels of gas
to migrate from regions of negative slope to those of positive slope.

The simple picture that emerges is that the WNM/CNM phase balance is driven
largely by the local ambient pressure. At ambient pressures typical of
galactic disks ($P/k$ of a few hundred to as much as 10000 ~cm$^{-3}$K) the
WNM and CNM can coexist in pressure equilibrium, while at even lower pressures
the WNM will dominate and at higher pressures it will be the CNM.

The timescale for equilibrium to be achieved has been determined by Draine
(\cite{draine78}) for quite a wide range of conditions to be about
$n\cdot\tau~\sim~10^6$ cm$^{-3}$yr. If ambient conditions (principally the
pressure) vary more rapidly than this, then equilibrium will not be
achieved and a broader range of conditions can be expected. In practice, one
can then expect a wider range of temperatures to be encountered at a given
location in a galaxy, than simply the two intercept points of a line of
constant pressure with the tilted ``S'' equilibrium curves. 

High resolution numerical simulations of the interstellar medium typically
show little evidence for a simple bimodal gas temperature distribution (eg. de
Avillez \& Breitschwerdt~\cite{deavillezand04}). This may reflect the
simplified treatment of the gas thermodynamics in the simulations, but may
also reflect the moderately short timescales for strong pressure modulation of
much of the medium.

However, there are some predicted global trends which can be sought in
observational data. A general decrease of the thermal pressure is expected as
a function of both radius, $R$, and $z$-height within disk galaxies. This
should give rise to a systematically changing phase balance as function of
both $R$ and $z$. At low $R$ and $z$ we expect the CNM to dominate, while at
high $R$ and $z$ it should be the WNM. Further, the pressure gradient as
function of $R$ might give rise to a systematic change in the median CNM
temperature.

In any case, the detailed thermodynamic analyses have shown that essentially
independent of the gas temperature, density and phase the neutral medium has
an associated electron density, $n_e~\sim~0.01$~cm$^{-3}$ (eg. Wolfire et
al. \cite {wolfireetal03}). This residual ionization is crucial in providing
coupling of the neutral medium to magnetic fields (which are after all the
focus of this conference).

\section{Observed Thermodynamics, Phase Balance and Temperature}

While the emission brightness of the 21~cm line of neutral hydrogen gives an
unambiguous indication that some atomic gas is present along a line-of-sight,
it does not always provide a direct constraint on the gas temperature or even
the total column density, in view of the unknown opacity. The best constraints
on physical conditions come from observations of both HI absorption and
emission from essentially the same line-of-sight. Such data have been
accumulated for both the solar neighborhood of the Galaxy and different
portions of the M31 disk by Braun \& Walterbos (\cite{braunand92}). The solar
neighborhood data show a very tight correlation of HI absorption opacity,
$\tau$, with emission brightness, $T_B$, that can be well-modeled with a two
phase ISM, although with about a factor of two variation in the CNM
temperature (100$\pm$50~K). The data for the M31 disk is much sparser and
noisier but shows a similar trend. These data further show a systematic trend
for the $\tau$--$T_B$ curves at increasing galacto-centric radii to be offset
to higher CNM temperatures.

\begin{figure}[ht]
 \resizebox{\hsize}{!}{ \includegraphics[keepaspectratio=true]{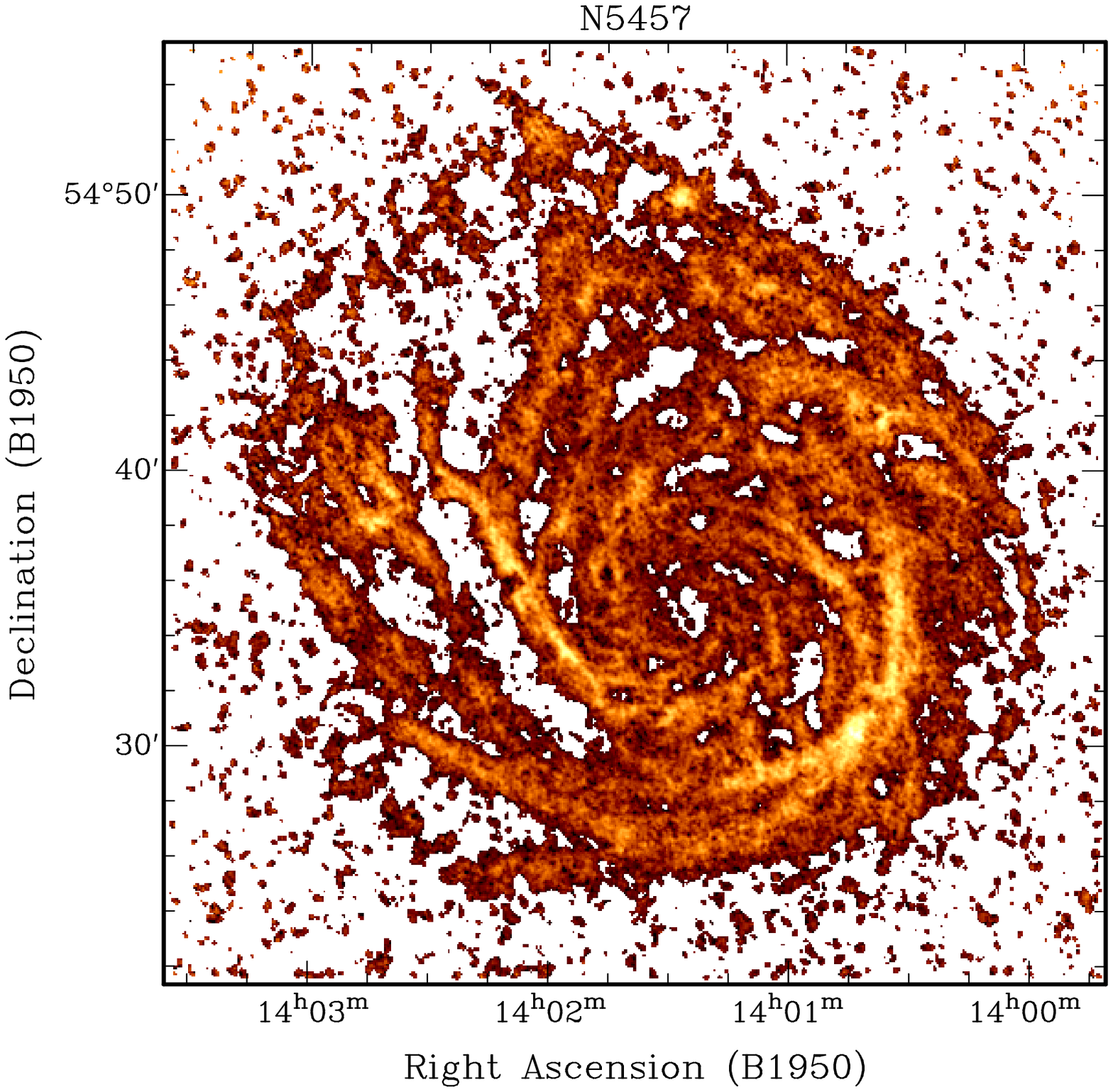} 
      \includegraphics[keepaspectratio=true]{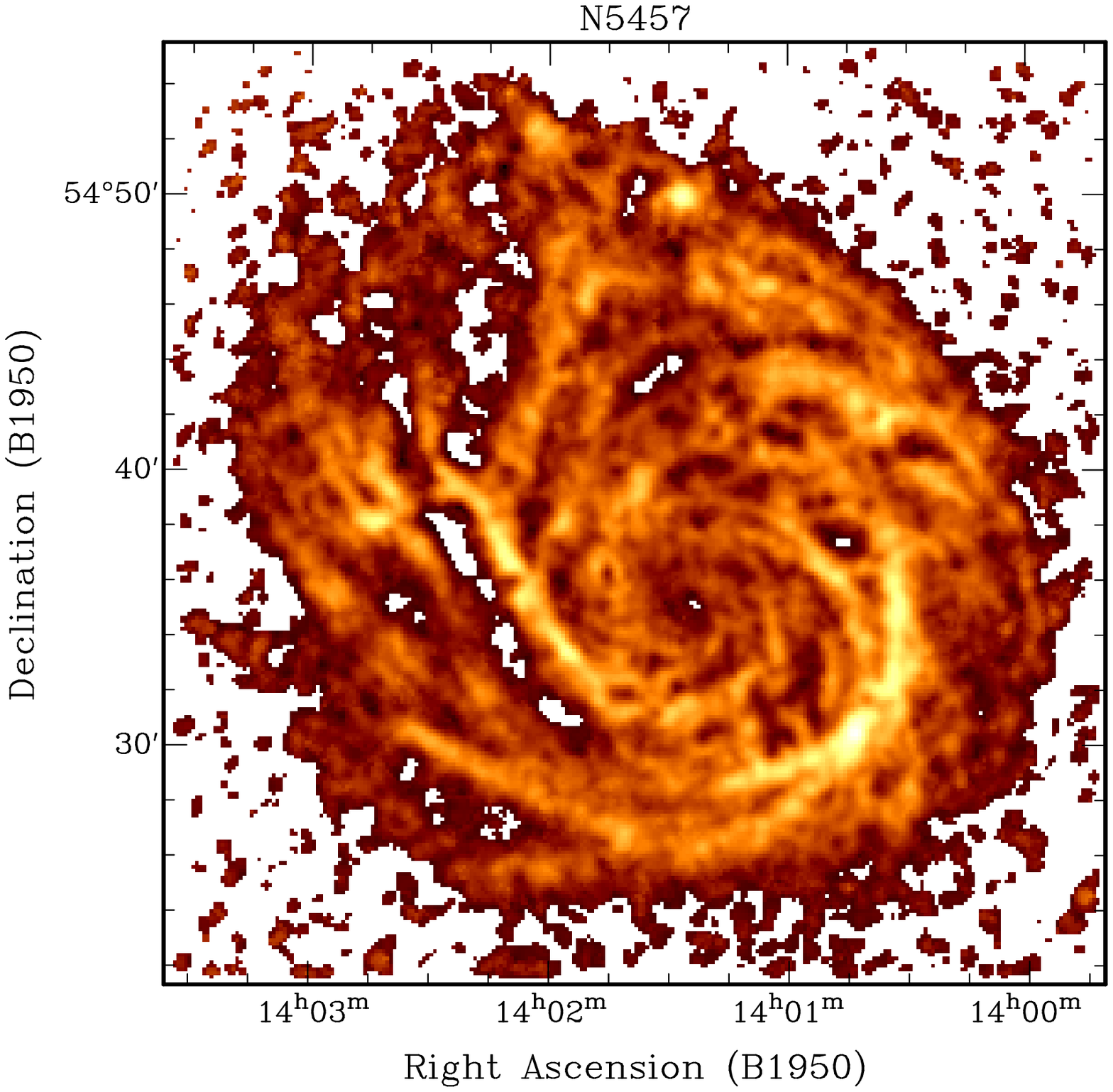}}
 \caption{Integrated HI emission in the nearly face-on galaxy NGC~5457. With
 high angular resolution {\bf (left)} a network of high brightness CNM
 filaments can be detected that is marginally resolved at 150~pc and 5
 km/s. Such filaments are confined to the star-forming disk of the galaxy,
 where they have a $\sim$15\% surface covering factor and account for more
 than half of the HI mass. With further spatial smoothing {\bf (right)}, much
 fainter diffuse emission is detected between the filaments and out to much
 larger radii consistent with the WNM phase.}
 \label{n5457cnmwnm}
\end{figure}

Similar conclusions follow from high resolution HI imaging of many nearby
galaxies by Braun~(\cite{braun95, braun97}). Such imaging studies reveal
a CNM network of high brightness temperature filaments that is marginally
resolved at about 150~pc and 5~km/s resolution in face-on systems. This is
superposed on a much fainter background of diffuse, higher line-width emission
that is consistent with arising within the WNM. These components are
contrasted in Fig.~\ref{n5457cnmwnm} for NGC~5457, seen at 9 arcsec resolution
on the left and after smoothing to 25 arcsec resolution on the right. When the
CNM has been marginally resolved it displays a peak observed brightness
temperature which increases systematically with galacto-centric radius from
some 50~K at small radii to about 200~K at the edge of the star-forming
disk. Several spot measurements of HI absorption toward background radio
sources confirm that this trend reflects an increase in the nominal CNM
kinetic temperature which is closely traced by the observed peak brightness
temperature. The implication of this correspondence is that the typical HI
opacity in the filaments is greater than unity. The surface covering factor of
the CNM filaments varies somewhat from galaxy to galaxy but is typically some
15\% inside of the star-forming disk, and essentially 0\% at larger
radii. These filaments account for 60--90\% of the total HI inside the
star-forming disk, but since the diffuse WNM continues to much larger radii,
the global CNM fraction is only some 35--70\%.

\begin{figure}[ht]
 \resizebox{\hsize}{!}{ \includegraphics[keepaspectratio=true]{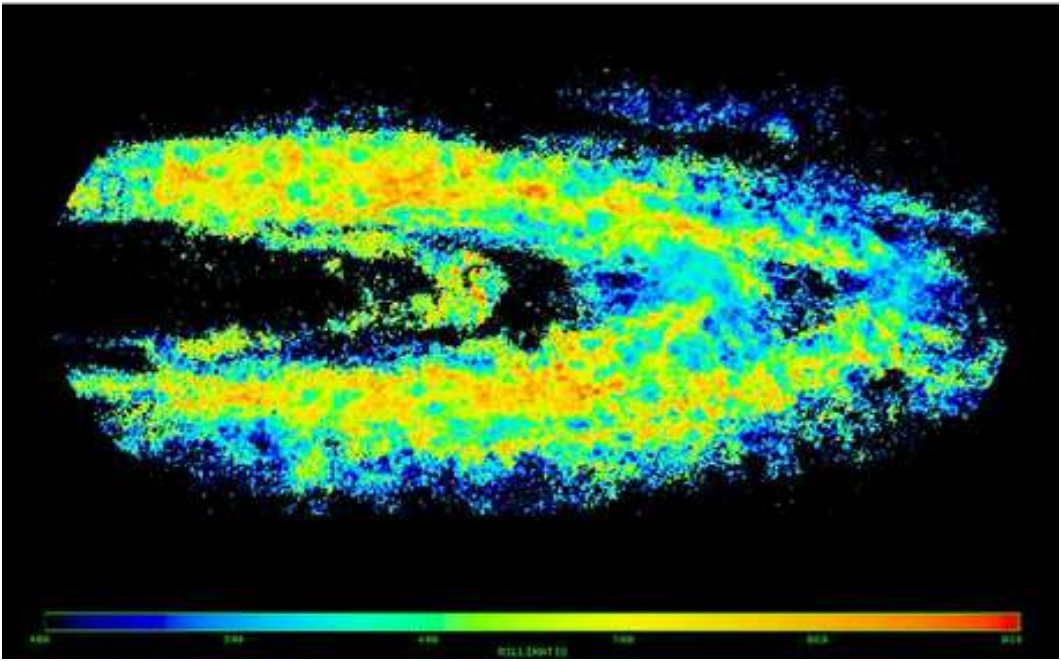}}
 \caption{The distribution of HI linewidth in the north-east half of
 M31. Regions of high linewidth form a filamentary network which
 appears to be organized and powered by massive star formation events.
 }
 \label{m31vel}
\end{figure}

Studies of the shape, rather than just the strength, of HI emission lines have
also provided insights into the role and origin of turbulence in the neutral
interstellar medium. The high brightness line profiles of CNM filaments are
found to be highly non-Gaussian in shape (Braun~\cite{braun97,
braun99}). Although the line cores are rather narrow, they are associated with
extremely broad wings ($\sim$100 km/s FWZI), which can be well-fit with
Lorentzian distributions. Just as the CNM itself is organized into a highly
filamentary system, so it is also the case with the regions of high HI
line-width. This phenomenon is clearly seen in Fig.~\ref{m31vel} for the
north-east half of M31. Even though the inclination of the M31 disk is so high
(almost 80$^\circ$) that the CNM filaments become highly blended when seen in
total intensity, they are readily distinguished in this image of HI 
line-width. Regions of high HI line-width are organized into a frothy system of
bubbles and filaments on scales of 10~pc up to about 1~kpc. Many of the
shell-like structures are associated with recent sites of massive star
formation as traced by H$\alpha$ emission and broad-band optical detection of
stellar associations. 

\section{Predictions for Galaxy Disks, Halos and Environs}

Quantitative predictions of neutral medium structures and topologies within
galaxy disks and halos are only beginning to emerge from recent high
resolution 3-D simulations (eg. de Avillez \&
Breitschwerdt~\cite{deavillezand04}) of $\sim$1~kpc$^2$ regions. Up to this
point, most predictions had been fairly qualitative and been based primarily
on the so-called ``Galactic Fountain'' (Shapiro \& Field~\cite{shapiroand76})
and ``chimney'' models (Norman \& Ikeuchi~\cite{normanand89}). Even now, it
has not been possible to simulate a complete realistic galaxy disk and halo in
three dimensions. This is becoming increasingly urgent as widespread
detections of halo gas with peculiar kinematics are being made (as we will see
below).

Various authors (eg. Maloney~\cite{maloney93}, Corbelli \&
Salpeter~\cite{corbelliand94}, Dove \& Shull~\cite{doveand94}) have considered
what should occur at the edges of galaxy disks, where the intergalactic
radiation field leads to a rapid decline in the neutral fraction. Below a
critical column density of log(N$_{HI}$)~$\sim$~19.5, there should be an
exponential decline in the neutral fraction from essentially unity down to
about 3\% by log(N$_{HI}$)~$\sim$~18. This marks the transition between the
optically thick and optically thin regime for the penetration of ionizing
intergalactic radiation. Ironically, the neutral fraction remains almost
constant at even lower column densities, once the optically thin regime has
been reached. However, in many ways this transition marks the effective
``edge'' of disk galaxies in HI, since current observations only rarely have
the sensitivity to reach column densities below log(N$_{HI}$)~$\sim$~19.

\begin{figure}[ht]
 \resizebox{\hsize}{!}{ \includegraphics[keepaspectratio=true, width=7cm]{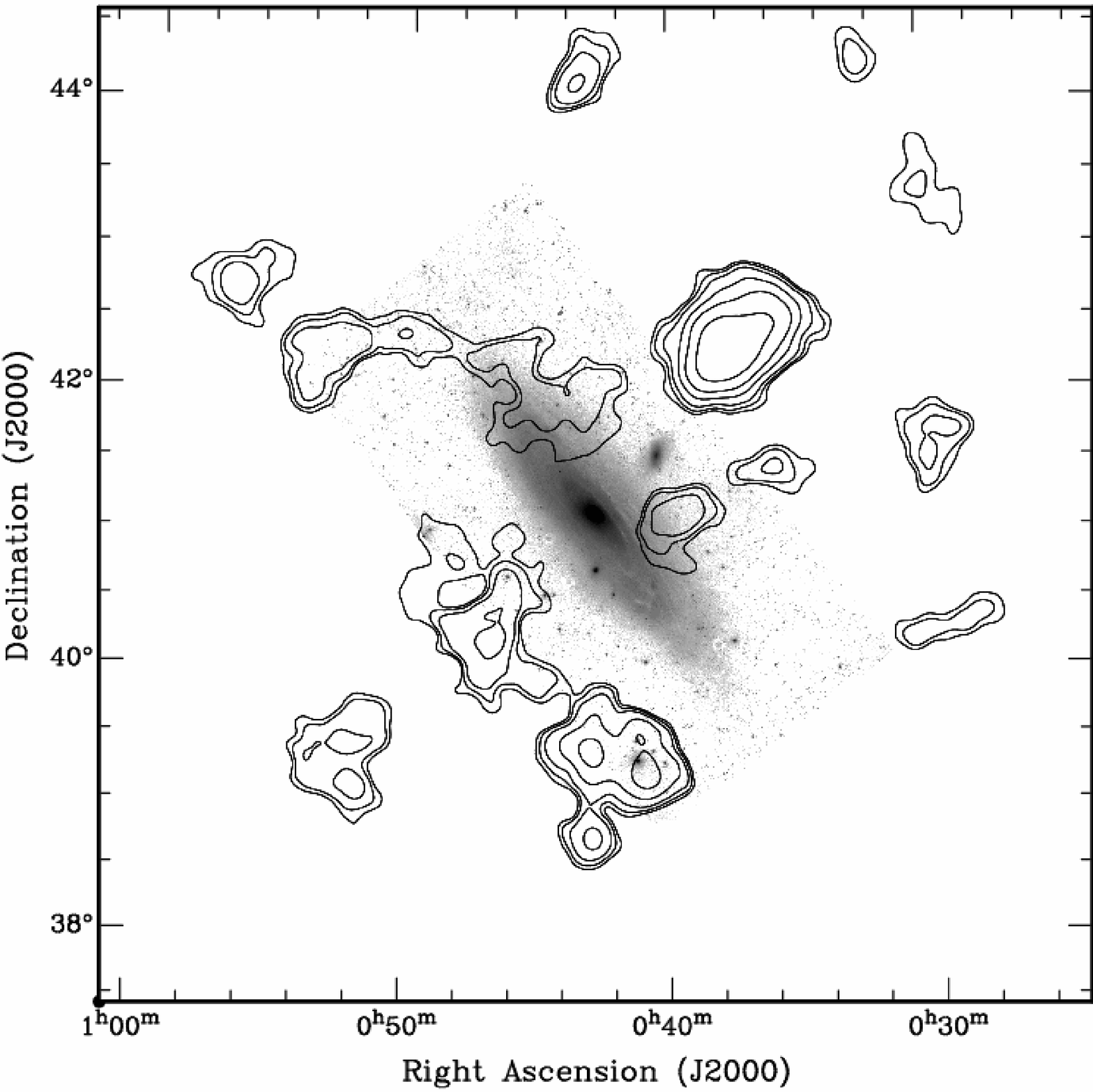} 
      \includegraphics[keepaspectratio=true, width=6cm]{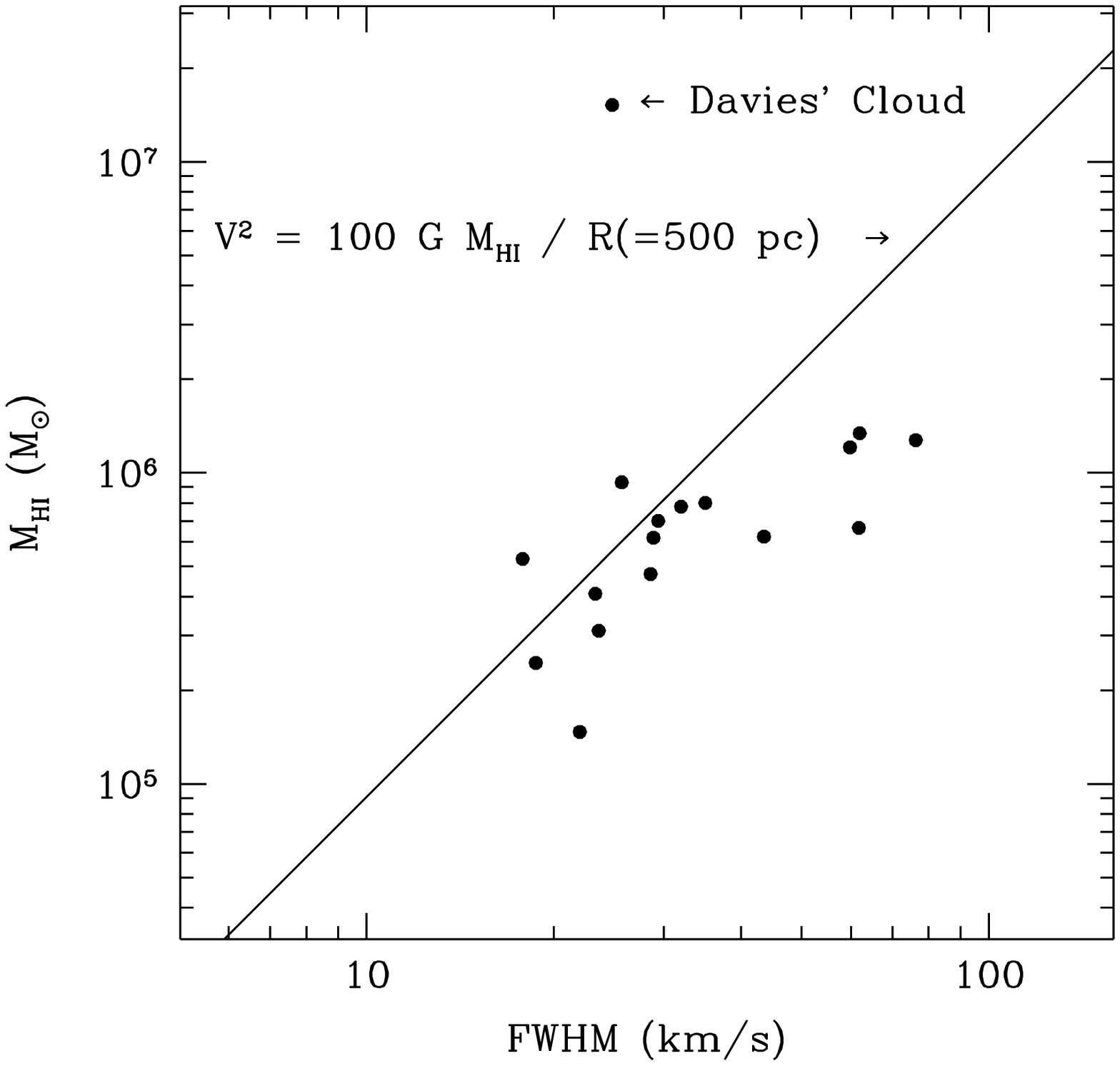}}
 \caption{Contours of HI column density {\bf (left)} for discrete and
diffuse high-velocity \ion{H}{1} in the M31 GBT field, after masking out
emission from the inclined, rotating disk; at 0.5, 1, 2, 10, and
$20\times10^{18}$~cm$^{-2}$, overlaid on a V band image of M31.  Observed
\ion{H}{1} mass versus FWHM line-width {\bf (right)} for discrete clouds near
M31.  Line-width was measured globally for each cloud.  Also plotted is
the anticipated relationship between \ion{H}{1} mass and line-width for
gravitationally confined clouds.  To first order, our FWHM measurements appear
consistent with the hypothesis that the objects are dark matter dominated,
assuming a dark matter to \ion{H}{1} mass ratio of 100:1 and a characteristic
size of 0.5 kpc for each \ion{H}{1} cloud core.  }
 \label{m31hvc}
\end{figure}

But galaxy disks, typically associated with log(N$_{HI}$)~$\sim$~18--22, only
account for about 1/3 of the cosmic baryons in the nearby universe. A further
third is apparently associated with more diffuse systems of ongoing
filamentary accretion in the range log(N$_{HI}$)~$\sim$~14--18. The final
third of local baryons likely resides in a very diffuse photo-ionized phase
and is associated with log(N$_{HI}$)~$\sim$~12--14. This breakdown in three
column density regimes is determined from high resolution numerical
simulations of structure formation over cosmic time (eg. Dav\'e et
al.~\cite{daveetal99, daveetal01}). These simulations suggest that in the
intermediate regime of log(N$_{HI}$), a large fraction of the gas resides in
the so-called ``warm-hot intergalactic medium'' or WHIM, a condensed
shock-heated phase with temperature in the range, T~$\sim~10^5$--$10^7$~K.
This simple picture is complicated by the growing suspicion that the gas
accretion process may well occur in two rather different regimes
(eg. Binney~\cite{binney04}, Keres et al.~\cite{keresetal04}). Low to moderate
mass galaxies (M$_{Vir}~<~10^{12}$M$_\odot$) may experience primarily
``cold-mode'' accretion (T~$\sim$~10$^{4.5}$~K) along filaments, while only
more massive systems may be dominated by the more isotropic ``hot-mode''
accretion (T~$\sim$~10$^{5.5}$~K).

\begin{figure}[ht]
 \resizebox{\hsize}{!}{ \includegraphics[keepaspectratio=true, width=7cm]{braun_fig4a.ps} 
      \includegraphics[keepaspectratio=true, width=6cm]{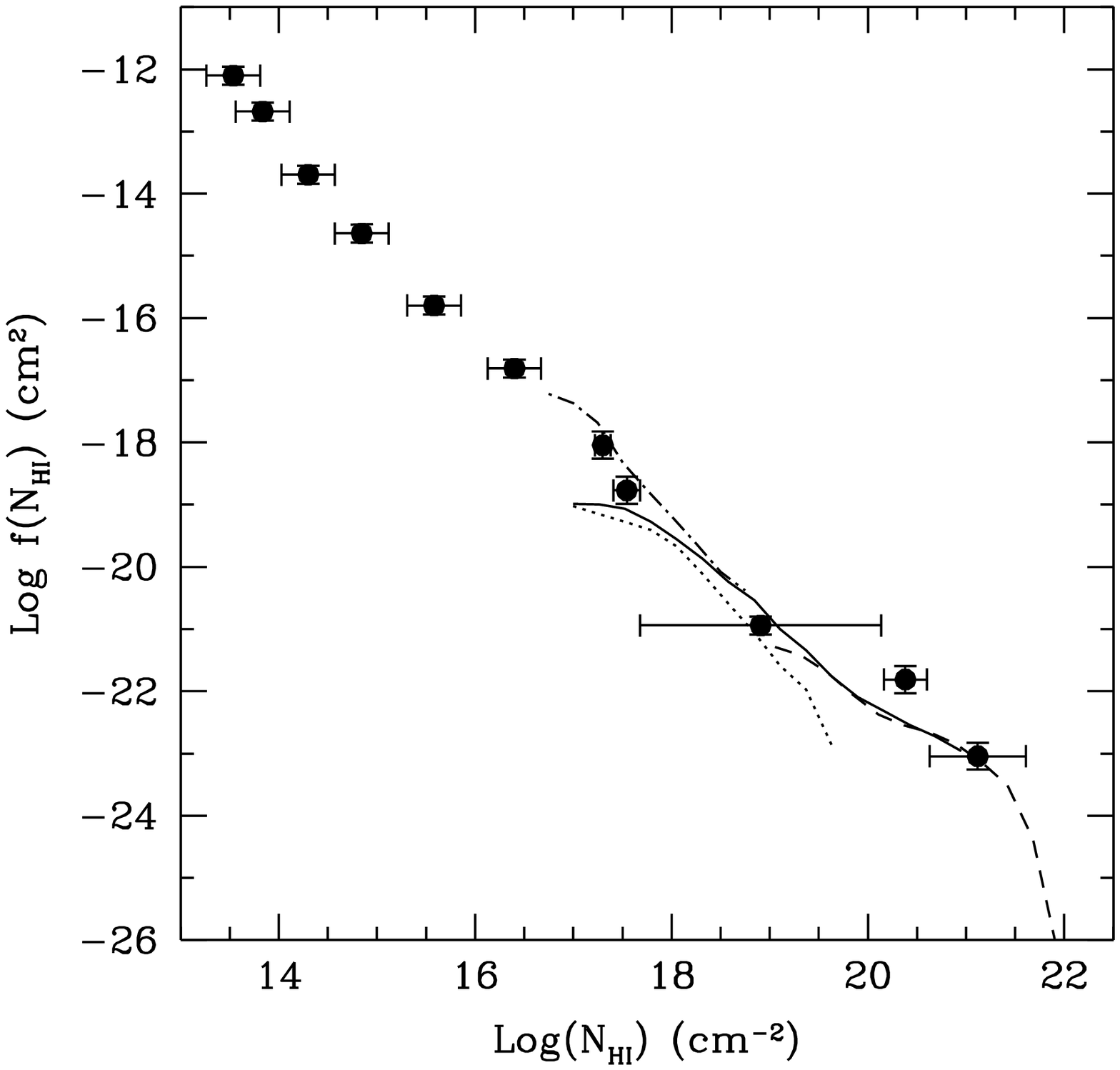}}
 \caption{Integrated HI emission {\bf (left)} from features which are
 kinematically associated with M31 and M33.  The grey-scale varies between
 log(N$_{HI}$)~=~17~--~18, for N$_{HI}$ in units of cm$^{-2}$. Contours are
 drawn at log(N$_{HI}$)~=~17, 17.5, 18, $\dots$ 20.5. M31 is located at
 (RA,Dec)~=~(00:43,+41$^\circ$) and M33 at (RA,Dec)~=~(01:34,+30$^\circ$), The
 two galaxies are connected by a diffuse filament joining the systemic
 velocities. The distribution function of HI column density {\bf (right)} due
 to M31 and it's environment. The data from three HI surveys of M31 are
 combined in this figure to probe column densities over a total range of some
 five orders of magnitude. The dashed line is from the WSRT mosaic (Braun et
 al. \cite{braunetal05}) with 1$^\prime$ resolution over 80$\times$40~kpc, the
 dotted and solid lines from our GBT survey (Thilker et
 al. \cite{thilkeretal04}) with 9$^\prime$ resolution over 95$\times$95~kpc
 and the dot-dash line from the wide-field WSRT survey (Braun \& Thilker
 \cite{braunand04}) with 48$^\prime$ resolution out to 150~kpc radius. The
 filled circles with error-bars are the low red-shift QSO absorption line data
 as tabulated by Corbelli \& Bandiera (\cite{corbelliand02}).}
 \label{m31m33}
\end{figure}

\section{Observations of Galaxy Disks, Halos and Environs}

A series of very deep studies of the HI distribution in galaxies of various
inclinations is beginning to reveal widespread evidence for components with
peculiar kinematics at large $z$-heights. One of the first
of these studies was that of NGC~2403 by Fraternali et
al.~\cite{fraternalietal02}. But more recently NGC~891 (Fraternali et
al.~\cite{fraternalietal05}), NGC~253 (Boomsma et al.~\cite{boomsmaetal05a})
and NGC~6946 (Boomsma et al.~\cite{boomsmaetal05b}) have all been subjected to
very sensitive imaging. What these studies are revealing is that
between 5 and 10\% of the total HI mass in the studied galaxies is distributed
in an extended ``halo'' component with an exponential scale-height of a few
kpc. In the most extreme cases, the gas has been detected as much as 15~kpc
out of the mid-plane. This high-$z$ gas rotates systematically more slowly
with height from the plane by some 20 to 50~km~s$^{-1}$. There is also a
component of systematic radial inflow of 15 to 20~km~s$^{-1}$ for this gas. In
relatively face-on systems it has been possible to detect localized vertical
outflows as large as 200~km~s$^{-1}$. This last attribute in particular can be
quite confidently associated with a ``Galactic Fountain'' or ``chimney''
phenomenon. It seems likely that other aspects of the halo gas are also
associated with these phenomena. 

However, other mechanisms which may well be important in fueling galaxy halos
are (1) the tidal disruption of satellite galaxies and (2) accretion of both
diffuse and discrete structures from the extended environment. A recent deep
study of a 95$\times$95~kpc field centered on M31 has been carried out with
the Green Bank Telescope (GBT) by Thilker et al. (\cite{thilkeretal04}). This
survey reaches an RMS sensitivity of 1.5$\times10^{17}$cm$^{-2}$ over
18~km~s$^{-1}$ and detects a multitude of peculiar velocity HI clouds and
streams, extending over the entire surveyed region, as shown in
Fig.~\ref{m31hvc}. Several of the detected features, particularly the complex
to the south of the M31 disk, are very likely tidal in origin since they
partially overlap with clear stellar streams in position and velocity
(eg. Ferguson et al.~\cite{fergusonetal02}). Other features in the M31
environment are less likely to have a tidal origin, since they have no stellar
counterparts, are well-separated from the disk, yet have large internal
velocity widths, which are rather suggestive of a significant dark matter
content. As seen in the right-hand panel of Fig.~\ref{m31hvc}, the
mass/line-width distribution of discrete clouds in the M31 environment is
suggestive of a typical dark (and/or ionized) to HI mass ratio of about 100:1.

On even larger scales, there is evidence for large-scale accretion of gas
along ``cosmic web'' filaments providing continued fueling of both M31 and
M33. Braun \& Thilker (\cite{braunand04}) have used the fourteen telescopes of
the WSRT array to image an 1800 deg$^2$ region centered on M31 with
auto-correlation spectra. This survey reaches an RMS column density
sensitivity of 4$\times10^{16}$cm$^{-2}$ over 17~km~s$^{-1}$, making it the
deepest HI look yet at the nearby extragalactic sky. In addition to detecting
many of the discrete features noted above within about 50~kpc of M31, a
diffuse filament is detected connecting the systemic velocities of M31 and M33
and extending beyond M31 in the anti-M33 direction, as seen in
Fig.~\ref{m31m33}. The total extent of this feature is some 260~kpc. In the
immediate vicinity of both M31 and M33 it terminates in what appear to be
ongoing accretion features. The distribution function of HI column density in
the M31 environment is also shown in Fig.~\ref{m31m33}, where it is compared
with the statistics from QSO absorption lines. The good agreement illustrates
that we are now making the first images of so-called Lyman Limit Systems,
which had previously only been detected via Ly$\alpha$ absorption. As noted in
the previous section, we expect that the HI at these low column densities is
the trace neutral fraction (about 1\%) in what is probably a highly ionized
plasma. Indeed, there are already some indications that much of the filament
shown in Fig.~\ref{m31m33} has a kinetic temperature of 2$\times10^5$ K (Braun
\& Thilker~\cite{braunand05}).


\begin{thebibliography}{}

\bibitem[2004]{binney04} 
Binney J., 2004, MNRAS, 347, 421
\bibitem[2005a]{boomsmaetal05a}
Boosma R., Oosterloo T., Fraternali F., van der Hulst, J.M., Sancisi R., 2005,
A\&A, in press, astro-ph/0410055 
\bibitem[2005b]{boomsmaetal05b}
Boosma R., Oosterloo T., Fraternali F., van der Hulst, J.M., Sancisi R., 2005,
in ``Extra-planar Gas'', Ed. R. Braun, ASP Vol. 331, in press,
astro-ph/0410022 
\bibitem[1992]{braunand92}
Braun R., Walterbos R.A.M., 1992, ApJ, 386, 120
\bibitem[1995]{braun95}
Braun R., 1995, A\&AS, 114, 409
\bibitem[1997]{braun97}
Braun R., 1997, ApJ, 484, 637
\bibitem[1999]{braun99}
Braun R., 1999, in ``Interstellar Turbulence'', Eds. J. Franco and
A. Carraminana, CUP, p.12
\bibitem[2004]{braunand04}
Braun R., Thilker D.A., 2004 A\&A, 417, 421
\bibitem[2005]{braunand05}
Braun R., Thilker D.A., 2005,
in ``Extra-planar Gas'', Ed. R. Braun, ASP Vol. 331, in press
\bibitem[2005]{braunetal05} 
Braun R., Thilker D.A., Corbelli E., Walterbos R.A.M., 2005, in prep.
\bibitem[1994]{corbelliand94}
Corbelli E., Salpeter E.E., 1994, ApJ, 419, 104
\bibitem[2002]{corbelliand02}
Corbelli E., Bandiera R., 2002, ApJ, 567, 712
\bibitem[1999]{daveetal99} 
Dav\'e R., Hernquist L., Katz N., Weinberg D.H. 1999, ApJ, 511, 521
\bibitem[2001]{daveetal01} 
Dav\'e R., Cen R., Ostriker J.P., et al. 2001, ApJ, 552, 473 
\bibitem[2004]{deavillezand04}
de Avillez M.A., Breitschwerdt D., 2004, A\&A, 425, 899
\bibitem[1994]{doveand94}
Dove J.B., Shull J.M., 1994, ApJ, 423, 196
\bibitem[1978]{draine78}
Draine B.T., 1978, ApJS, 36, 595
\bibitem[2002]{fergusonetal02}
Ferguson, A.M.N., Irwin M.J., Ibata R.A., Lewis G.F., Tanvir N.R., 2002, 
AJ, 124, 1452
\bibitem[1965]{field65}
Field G.B., 1965, ApJ, 142, 531
\bibitem[2002]{fraternalietal02}
Fraternali F., van Moorsel G., Sancisi R., Oosterloo T., 2002, AJ, 123, 3124
\bibitem[2005]{fraternalietal05}
Fraternali F., Oosterloo T., Sancisi R., Swaters, R., 2005, in ``Extra-planar
Gas'', Ed. R. Braun, ASP Vol. 331, in press, astro-ph/0410375
\bibitem[2004]{keresetal04} 
Keres D., Katz N., Weinberg D.H., Dav\'e R. 2004, MNRAS, astro-ph/0407095  
\bibitem[1993]{maloney93}
Maloney P., 1993, ApJ, 414, 41
\bibitem[1989]{normanand89}
Norman C.A., Ikeuchi S., 1989, ApJ, 345, 372
\bibitem[1976]{shapiroand76}
Shaprio P.R., Field G.B., 1976, ApJ, 205, 762
\bibitem[1985]{shulland85}
Shull J.M., Woods D.T., 1985, ApJ, 288, 50
\bibitem[2004]{thilkeretal04} 
Thilker, D.\,A., Braun, R., Walterbos, R.\,A.\,M., et al. 2004 ApJ, 601, L39
\bibitem[2003]{wolfireetal03}
Wolfire M.G., McKee C.F., Hollenbach D., Tielens A.G.G.M., 2003, ApJ, 587,
278 

\end{thebibliography}
\end{document}